\documentclass{aa}  
\usepackage{graphicx}
\usepackage{txfonts}
\usepackage{xcolor}

\begin{document}


\newcommand{\Msun}{\rm M_\odot}
\newcommand{\Mvir}{\rm M_{\rm vir}}
\newcommand{\Mlens}{\rm M_{\rm lens}}
\newcommand{\Mstel}{\rm M_{\rm stel}}
\newcommand{\rvir}{\rm r_{\rm vir}}
\newcommand{\cvir}{\rm c_{\rm vir}}
\newcommand{\rs}{r_s}
\newcommand{\re}{r_{\rm eff}}

\title{Reconciling concentration to virial mass relations}


\author{Dominik Leier\inst{1}\fnmsep\thanks{E-mail: dominik.leier@gmail.com}
  \and
  Ignacio Ferreras\inst{2}\fnmsep\inst{3}\fnmsep\inst{4}\fnmsep\thanks{E-mail: ignacio.ferreras@iac.es}
  \and
  Andrea Negri\inst{2}\fnmsep\inst{4}
  \and
  Prasenjit Saha\inst{5}}

\institute{Dipartimento di Fisica e Astronomia, Alma Mater Studiorum Universit\`{a} di Bologna, Viale B. Pichat 6/2, 40127, Bologna, Italy
\and
Instituto de Astrof{\'i}sica de Canarias, Calle V{\'i}a L{\'a}ctea s/n,
E38205, La Laguna, Tenerife, Spain
\and
Department of Physics and Astronomy, University College London, London WC1E 6BT, UK
\and
Departamento de Astrof\'\i sica, Universidad de La Laguna, E38206 La Laguna, Tenerife, Spain
\and
Physik-Institut, University of Z\"urich, Winterthurerstrasse 190, CH-8057 Z\"urich, Switzerland}

\date{Accepted for publication in Astronomy \& Astrophysics, October 2024}

\abstract{The concentration--virial mass (c-M) relation is a
  fundamental scaling relation within the standard cold dark matter ($\Lambda$CDM) framework well
  established in numerical simulations. However, observational
  constraints of this relation are hampered by the difficulty of
  characterising the properties of dark matter haloes. Recent comparisons
  between simulations and observations have suggested a systematic difference
  of the c-M relation, with higher concentrations in the latter.}
{In this work, we undertake detailed comparisons between simulated galaxies and
  observations of a sample of strong-lensing galaxies.}
{We explore several factors of the comparison with strong
  gravitational lensing constraints, including the choice of the
  generic dark matter density profile, the effect of radial resolution, the reconstruction limits of observed versus simulated mass
  profiles, and the role of the initial mass function in the
  derivation of the dark matter parameters. Furthermore, we show the
  dependence of the c-M relation on reconstruction and model errors
  through a detailed comparison of real and simulated gravitational
  lensing systems.}
{An effective reconciliation of simulated and observed c-M
  relations can be achieved if one considers less strict
  assumptions on the dark matter profile, for example, by changing the slope
  of a generic NFW profile or focusing on rather extreme combinations of stellar-to-dark
  matter distributions. A minor effect is inherent to the applied
  method: fits to the NFW profile on a less well-constrained inner mass profile yield
  slightly higher concentrations and lower virial masses.}
{}


\maketitle

\section{Introduction}

The standard paradigm of galaxy formation rests on the presence of two
main ingredients: dark matter and baryons. Both are subject to
gravitational forces, but only the latter can be detected with
photons. In this framework, dark matter represents a fundamental
component, as it provides the
backbone structure over large scales (cosmic web) and within the
scales in which galaxies are found (haloes). The properties of dark
matter haloes are thus essential to understanding galaxy formation, 
but these properties are limited to indirect constraints involving
observations of the baryonic matter. Gravitational lensing provides a
special method of detection by using the distortions exerted on the
photons emitted by a background source as they pass through a
gravitating structure. By `removing' the contribution of the baryons from the
lensing signal, it is thus possible to study dark matter haloes.

One of the main correlations found in haloes is the concentration versus
mass relation (hereafter c-M, \citealt{NFW}), by which more massive
haloes tend to have progressively lower concentrations, albeit with a
large scatter.  This relation has been readily found in simulations
\citep[e.g.][]{Bullock:01,Maccio:07,Dutton:14,Diemer:19,Ishiyama:21},
and its scatter could unravel the details concerning the growth of
haloes \citep[e.g.][]{Wang:20}. As a first approximation, this trend
is a direct consequence of the bottom-up scenario of structure
formation, as lower mass haloes formed statistically at earlier cosmic
times, when the density in the expanding Universe was higher
\citep[see, e.g.][]{MvdBW:10}. The concentration of dark matter haloes
depends on the adopted cosmology \citep{Maccio:08}. However, this
result is expected from a simple spherical collapse scenario, whereas
a more realistic depiction would involve non-spherical structures and
extended mass assembly histories, making the interpretation of
concentration more complicated. Virialised haloes, in principle,
should have higher concentrations \citep{Neto:07}. If the subsequent
growth after the formation of the "first" structure is slow, we can
expect a gradual increase of the concentration with total mass,
whereas faster growth can keep the concentration unchanged
\citep{Correa:15}. Therefore, variations among haloes on the c-M plane
depend on their mass assembly history.  The large scatter of the
c-M relation found in simulations is thus an indicator of the diverse
formation histories of structures. While virial mass is a good guess
for a first-order parameter, more information is needed to describe
the details of individual haloes (e.g. assembly bias;
\citealt{Wechsler:06}). For instance, at a fixed halo mass, one would
expect `older' haloes to populate the high concentration envelope
of the c-M relation. The dynamical state of a halo -- shape, spin,
virialisation -- also affects its location on the c-M plot.

Observational constraints are harder to come by and mostly rely on
the X-ray emission from the hot gas surrounding the most massive
structures \citep[e.g.][]{Buote:07}. Gravitational lensing offers a
complementary method to determine the properties of dark matter haloes
\citep[e.g.][]{Comerford:07,Mandelbaum:08,Merten:15}. Moreover, when
restricting the lenses to galaxy scales, it is possible to constrain
dark matter haloes for masses below $\lesssim 10^{13}$M$_\odot$
\citep[e.g.][]{Leier:11}.  However, all observational methods are
based on an indirect detection of the dark matter via the
gravitational potential.  Simulations such as EAGLE \citep{EAGLE} or
Illustris \citep{Illustris} do not suffer from this disadvantage, as
the dark matter distribution is directly accessible from the
simulation outputs.  However, even the simulations rely on the
parameterisation of mass and concentration, which require the adoption
of a specific density profile function, such as those proposed by
\citet{NFW} or \citet{Einasto:65}.

The goal of this work is to assess the concentrations derived from
strong lensing analysis over galaxy scales, which appear
inconsistently high with respect to the predictions from numerical
simulations \citep[e.g.][]{LFS12}. We show how the
extracted c-M values (see Table~\ref{tab:cm}) depend on resolution,
modelling uncertainties, and
assumptions of the underlying analytic functions, such as the dark
matter profile and the initial mass function (IMF). We build upon a recent
analysis of the c-M relation \citep{LFS22} in a sample of
strong-lensing galaxies versus simulated galaxies from EAGLE using a
Monte-Carlo type combination of pixelised lens models \citep{PIXELENS}
and stellar population synthesis maps, as used in \citet{FSW:05}.
In Sect.~\ref{sec:obsvssim} we describe the data, and
in Sect.~\ref{sec:cMlens}, the method used in this study is presented.

\begin{table}
\begin{center}
  \caption{Overview of the $c-\Mvir$ power-law parameters.}
\label{tab:cm}
\renewcommand{\arraystretch}{1.1} 
\begin{tabular}{@{}llcc@{}}
\hline
& Sample & $\alpha$ & $c_{13}$  \\
& \multicolumn{1}{l}{(1)} & \multicolumn{1}{c}{(2)} & \multicolumn{1}{c}{(3)} \\
\hline
\hline
& lit, B07   & $-0.199 \pm 0.026$ &  $14.42 \pm 0.91$ \\ 
& lit, L12   & $-0.401 \pm 0.064$ &  $17.70 \pm 3.89$ \\
\hline
& $lens,all$  & $-0.65 \pm 0.10$ &  $14.56^{+5.54}_{-4.02}$ \\
& $lens,<50$ & $-0.41 \pm 0.07$ &  $16.73^{+2.98}_{-3.62}$ \\
\hline
& $EAGLE,all$ & $-0.13 \pm 0.03$ &  $11.02^{+0.46}_{-0.49}$\\
& $EAGLE,{K13}$ & $-0.19 \pm 0.05$ & $8.18^{+0.57}_{-0.60}$ \\
\hline
\end{tabular}
\tablefoot{In col.~1, B07 refers to the results of \citet{Buote:07} and L12 to \citet{LFS12}. The samples ``lens''
    and ``EAGLE'' give the best fits from \citet{LFS22}. ``lens$_{<50}$'' denotes fits
    to our lens sample with a root mean square deviation better than the median value.
    Col.~2 is the slope of the scaling relation and col.~3 is the concentration of the best fit at virial mass $10^{13}$M$_\odot$. We provide errors from bootstrap.}
\end{center}
\end{table}

Our results in Sect.~\ref{sec:param} show how fitted c-M relations
change if the reconstructed lens profile is less well resolved in the
inner part -- in other words how the c-M relation changes as a function of the
minimum radius. Furthermore, we demonstrate how a transformation
that keeps the total enclosed mass at the Einstein radius constant
but changes the steepness of the inner profile affects the c-M
relation as well as the goodness-of-fit. In addition, we compare our
findings with non-parametric concentrations derived from the inner
radial region of lenses and simulated haloes alike. In
Sect.~\ref{ssec:emp}, we show how dark matter distributions based on
parametric functions other than the standard NFW profile \citep[named after][]{NFW}, change the c-M
relation. Subsequently, in Sect.~\ref{ssec:imf}, we study how
the adoption of different choices of the stellar IMF affects the c-M relation. Finally, in
Sect.~\ref{sec:disc}, we discuss whether or not all of these factors can
reconcile the observed and simulated c-M relations.

\section{Observed and simulated lenses}
\label{sec:obsvssim}
The lensing data consist of 18 systems selected to have 
a geometry that allows for the determination of the total mass from one to 
several effective radii of the lens galaxy. The sample has Hubble Space Telescope (HST) imaging in 
the optical (WFPC2 and ACS) and NIR (NICMOS)\footnote{see \url{https://www.stsci.edu/hst/instrumentation}
 for details of instrument acronyms}, which allowed us to produce  
a projected stellar mass map by combining the photometry with stellar population
synthesis models. This map is contrasted with a total mass map obtained 
from the lensing properties of the systems, adopting PixeLens 
\citep{PIXELENS} as the method to solve the lensing equations. 
We refer the reader to Table~\ref{tab:lens_data} and \citet{LFS16} for full details about the sample.
We note that the derivation of stellar mass includes the possibility of a non-standard stellar IMF. We adopt two choices: the so-called bimodal IMF (BM) and a two-segment power law (2PL). More details can
be found in \citet{LFS16}.

We retrieved the simulation data from EAGLE \citep{EAGLE}, more 
specifically the  z=0.1 snapshot of the RefL0100N1504 run. 
This simulation adopts a comoving box size of 100\,Mpc with 
$1054^3$ dark matter particles and also includes a gas or stellar mass
baryon component following standard hydrodynamical (SPH) modelling 
along with a set of subgrid prescriptions for the evolution of the
stellar, gas, and SMBH components \citep{Crain:15}.
From this sample, we selected 
all haloes harbouring galaxies with a stellar mass above $\log M_s/M_\odot=10.75$, 
thus representing systems comparable with the observed lenses.

As our aim is to address observed and simulated lenses with the same method, we 
produced projections of the mass distribution of EAGLE data in the same format
as those retrieved from {\sc PixeLens} when constraining the real data. 
Moreover, to account for the expected variance in the haloes due to orientation,
shape, and other properties, we drew an ensemble by randomly projecting the EAGLE systems (both 
stellar and dark matter particles) onto 100 random orientations. We then 
adopted the same lens geometry as the observed set  -- randomly assigning the 
parameters of one of the 18 lenses to the EAGLE data -- and took the same spatial
binning for full consistency. We note that we neglect the contribution of gas to
the mass budget of these galaxies, a choice justified by the (high) stellar mass of the
systems we targeted. In fact, the distribution of the gas to 
total baryon mass fraction within a projected radial distance of 15\,kpc has a mean and 
standard deviation of $0.039\pm 0.022$, justifying this approximation.

\begin{table*}
\begin{center}
\caption{Lens data overview.}
\label{tab:lens_data}
\renewcommand{\arraystretch}{1.1} 
\begin{tabular}{@{}rcccccccccl@{}}
\hline
\multicolumn{1}{c}{Lens ID}  & \multicolumn{1}{c}{$r_{lens}$}  &  \multicolumn{1}{c}{$r_{rec}$} & \multicolumn{1}{c}{$r_{lens}/r_{e,I}$} & \multicolumn{1}{c}{$\Sigma_{\text{crit}}$}  & \multicolumn{1}{c}{$z_{L}$} & \multicolumn{1}{c}{$c$} & \multicolumn{1}{c}{$R_{90}/R_{50}$} \\
& (") & (") & & $( \frac{10^{10} M_{\odot}}{\text{kpc}^2})$ & & & \\
\multicolumn{1}{c}{(1)} & \multicolumn{1}{c}{(2)} & \multicolumn{1}{c}{(3)} & \multicolumn{1}{c}{(4)} & \multicolumn{1}{c}{(5)} & \multicolumn{1}{c}{(6)} & \multicolumn{1}{c}{(7)} & \multicolumn{1}{c}{(8)} \\
\hline
\hline
J0037 & 1.309 & 4.06 & 0.73 & 3.92 & 0.1954 & $\phantom{0}9.16^{+23.8}_{-7.7}$  & $1.40^{+0.41}_{-0.05}$ \\
J0044 & 0.601 & 2.55 & 0.31 & 4.52 & 0.120 & $34.53^{+11.5}_{-12.3}$  & $1.55^{+0.09}_{-0.10}$ \\
J0946 & 1.541 & 3.08 & 0.66 & 2.83 & 0.222 & $21.48^{+12.6}_{-15.6}$ & $1.58^{+0.14}_{-0.19}$ \\
J0955 & 1.141 & 2.69 & 0.78 & 2.54 & 0.111 & $\phantom{0}8.12^{+30.1}_{-3.8}$ & $1.35^{+0.17}_{-0.01}$ \\
J0959 & 0.920 & 2.73 & 0.71 & 2.42 & 0.126 & $23.11^{+40.7}_{-19.6}$ & $1.45^{+0.43}_{-0.10}$ \\
J1100 & 1.341 & 3.65 & 0.61 & 6.39 & 0.317 & $\phantom{0}5.28^{+13.5}_{-3.2}$ & $1.39^{+0.23}_{-0.04}$ \\
J1143 & 1.104 & 4.75 & 0.41 & 2.15 & 0.106 & $21.32^{+31.3}_{-18.2}$ & $1.49^{+0.35}_{-0.14}$ \\
J1204 & 1.258 & 3.26 & 1.15 & 3.15 & 0.164 & $23.81^{+37.5}_{-18.4}$ & $1.51^{+0.37}_{-0.15}$ \\
J1213 & 1.115 & 3.65 & 0.74 & 2.24 & 0.123 & $18.23^{+39.0}_{-16.0}$ & $1.47^{+0.56}_{-0.12}$ \\
J1402 & 1.344 & 2.89 & 0.59 & 4.91 & 0.205 & $43.97^{*}_{-41.9}$ & $2.00^{+12.1}_{-0.65}$ \\
J1525 & 1.219 & 3.87 & 0.50 & 8.81 & 0.358 & $\phantom{0}5.30^{+9.98}_{-4.1}$ & $1.39^{+0.20}_{-0.04}$ \\
J1531 & 1.543 & 4.04 & 0.78 & 2.18 & 0.160 & $\phantom{0}4.06^{+11.0}_{-2.3}$ & $1.36^{+0.12}_{-0.01}$ \\
J1538 & 1.011 & 2.08 & 1.01 & 2.83 & 0.143 & $17.71^{+387.4}_{-15.5}$ & $1.47^{+4.48}_{-0.12}$ \\
J1630 & 1.687 & 4.63 & 1.02 & 4.82 & 0.248 & $\phantom{0}6.81^{+25.8}_{-5.8}$ & $1.39^{+0.38}_{-0.04}$ \\
J1719 & 1.228 & 2.62 & 0.84 & 3.70 & 0.181 & $19.31^{+655.2}_{-17.7}$ & $1.50^{+7.65}_{-0.15}$ \\
J2303 & 1.346 & 4.24 & 0.46 & 3.18 & 0.1553 & $10.71^{+20.9}_{-8.6}$ & $1.41^{+0.28}_{-0.06}$ \\
J2343 & 1.293 & 3.41 & -- & 4.16 & 0.181 & $19.46^{+48.3}_{-17.5}$ & $1.51^{+0.73}_{-0.17}$ \\
J2347 & 0.742 & 3.40 & 0.55 & 11.80 & 0.417 & $\phantom{0}8.95^{+13.3}_{-7.2}$ & $1.41^{+0.28}_{-0.06}$ \\
\vspace{-2mm} \\
\hline
\end{tabular}
\tablefoot{Col.~1 lists the lens IDs. Col.~2 shows the lens radius $r_{lens}$, an average radial measure of the position of lensed images, where the reconstruction uncertainty is minimised. Col.~3 lists the reconstruction radius. Col.~4 gives the ratio of $r_{lens}$ to the effective radius in the I band. Col.~5 shows the critical surface density $\Sigma_{\text{crit}}$. Col.~6 shows the redshift $z_{L}$ of the lens galaxy. Col.~7 and Col.~8 contain the parametric and non-parametric estimates of concentration shown in Fig.~\ref{fig:ccrr}.}
\end{center}
\end{table*}

\begin{figure}
\centering 
\includegraphics[width=0.9\columnwidth]{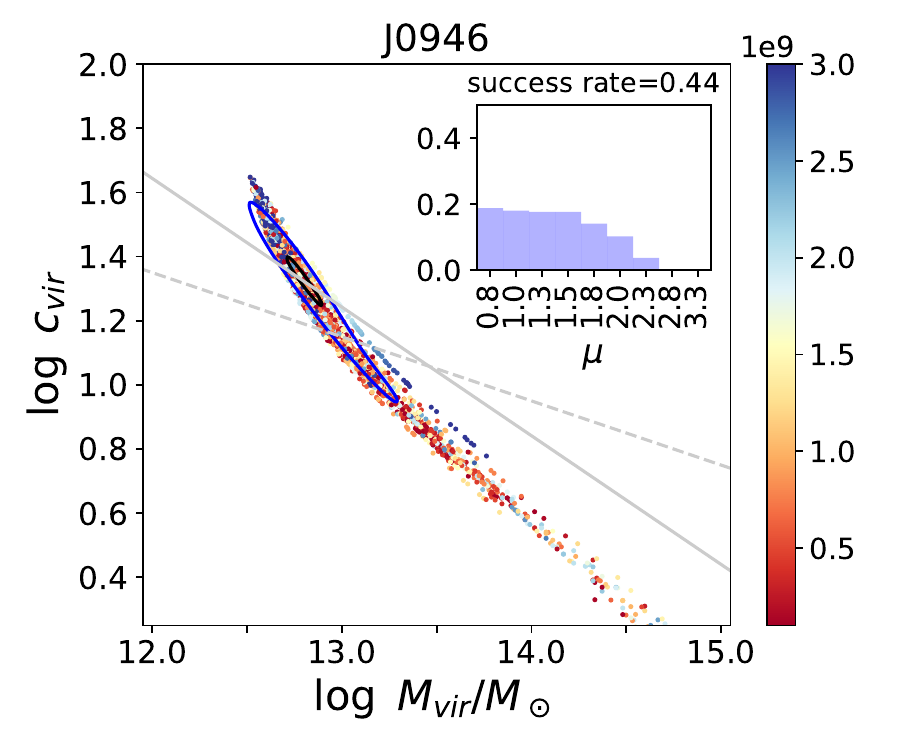}
\caption{Point cloud of successful combinations of lensing and stellar mass profiles on the c-M plane for lens J0946 representing the success rate. The colour scale indicates the MAE of the fitted models, where lower values correspond to a better goodness-of-fit. The $95\%$ ($68\%$) kernel density estimate contours are shown in blue (black). The solid grey line represents the c-M relation from \citet{LFS22}, while the dashed line is from \citet{Buote:07}. The inset displays the frequency distribution of successful realisations as a function of the IMF slope, with very bottom-heavy IMFs leading to dark matter profiles that result in unsuccessful fits.}
\label{fig:cm1}
\end{figure}

\section{Lensing constraints on the c-M plane}
\label{sec:cMlens}

The basic method of comparison is illustrated in Fig.~\ref{fig:cm1}
(applying the method to lens J0946 with the adoption of a 2PL IMF).
The concentration of a dark matter halo is a manifestly difficult measurement 
in observations of gravitating structures. To the lowest order, one can 
assess the mass to light ratio 
($\Upsilon\equiv M/L$) through dynamical or lensing measurements within the extent of the observations. Traditionally,
these constraints on $\Upsilon$ represent evidence of dark matter
in galaxies and clusters. At the next order of reliability, one can push the data towards regions where the baryon content is negligible (usually well beyond $2.5~\re$, according to \cite{Leier:11}) and come up with a measurement of the total halo mass. This depends on the assumptions made to determine the extent of the halo and on the accuracy of extrapolating the
observational constraints to the adopted limit of the halo.
Through the use of generic models for the
density profile of dark matter, one can determine a number of parameters.
If the measurements extend far enough from the baryonic structure, a reliable estimate of the halo mass can be obtained. A rough approximation of the physical scales for both the luminous and total mass can be inferred from Columns~2 and 4 of Table~\ref{tab:lens_data}.
This is also the case in 
simulations, where either tracing the 
gradient of the density profile or determining the distribution of bound particles 
allows one to find the limits of the halo.
The standard framework of structure formation identifies total halo mass as
the fundamental parameter. Indeed authors of work targeting environment-related properties
in galaxies \citep[e.g.][]{Rogers:10,Peng:12,AP:15} have chosen
halo mass as the driving factor to assess the interplay
between galaxy formation and the underlying dark matter
(any other dependence on halo parameters is loosely termed assembly bias).

The next order regarding the difficulty of derivation is the slope of the 
density profile. This involves a differentiation process that is 
much more prone to noise from variations in the distribution. To begin with,
one needs to assume a radial structure (spherical, ellipsoidal, etc.), which is
not necessarily a good representation in all cases, especially given the
collisionless nature of dark matter particles and the long relaxation times
of dark matter haloes. From this parameterisation, we derived an estimate of
concentration, defined as the ratio between the virial 
radius (or variations thereof) and the scale radius, given by the position
where the logarithmic slope is isothermal (i.e. $d\log\rho/d\log r=-2$).
This is where the observational approach struggles. While a total mass estimate
can be made without adopting many assumptions, the derivation of concentration 
is much more prone to the details. 

In the standard methodology based on the NFW profile, the c-M relation can also 
be considered as a correlation between the scale radius and the virial radius.
This paper focuses on how strong gravitational lensing estimates of the c-M relation
can be reconciled with the theoretical expectations from numerical simulations 
of structure formation. In addition to the methodology described in the previous section, 
we include in this paper the effect of variations in the stellar IMF.
The IMF is the distribution of stellar mass at birth in star forming regions. 
It is often treated as a universal function, based on constraints in the resolved stellar
populations of the Milky Way, and usually described by either a power 
law \citep{Salp:55} or variations that include a tapered low-mass end 
and differing slopes at the massive end \citep[see, e.g.][]{Kroupa:01,Chab:03}. 
Over the past decade or so, evidence has shown that the IMF can depart
significantly from the canonical one, especially in the dense central 
regions of early type galaxies \citep[see, e.g.][]{IMN:15,FLB:19}.

In our analysis, the effect of a non-universal IMF has to do with the
derivation of the stellar mass profile and hence the dark matter halo --
produced as the remainder from the lensing mass distribution. In gravitational
lensing systems, the IMF has proven controversial 
\citep[see, e.g.,][]{Smith:15,Smith:17,LFS16}, showing results at odds with
the alternative constraints based on stellar population analysis 
\citep[e.g.][]{vdC:10,IMF:13,FLB:19} or 
dynamics \citep[e.g.][]{Capp:12,Lasker:13,Lyub:16}. In this paper, we only consider the effect that 
a non-universal IMF might have in a systematic variation on the c-M plane
determined for strong lenses.

\section{Resolution and non-parametric concentrations}
\label{sec:param}

The focus of this paper is to explore in more detail the offset found in
\citet{LFS22} between observed galaxies (via gravitational lensing) and
simulated galaxies. While both samples show the characteristic negative
correlation between concentration and halo mass, the lensed galaxies 
suggest higher values at a fixed mass with respect to the simulations.
Needless to say, the derivation
of halo parameters from the observations is dependent on a number of
assumptions, which we study in the following, that could affect the systematic
bias in the comparison made on the concentration versus mass relation plane.

\subsection{Inner profile}
\label{ssec:inner}
\begin{figure}
\centering 
\includegraphics[scale=0.27]{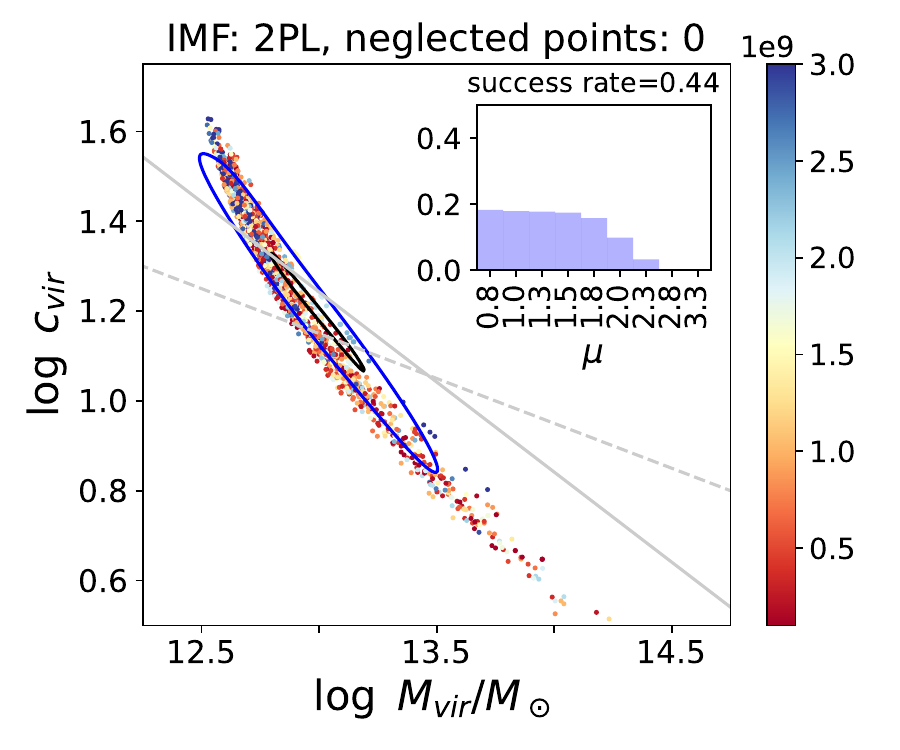}
\includegraphics[scale=0.27]{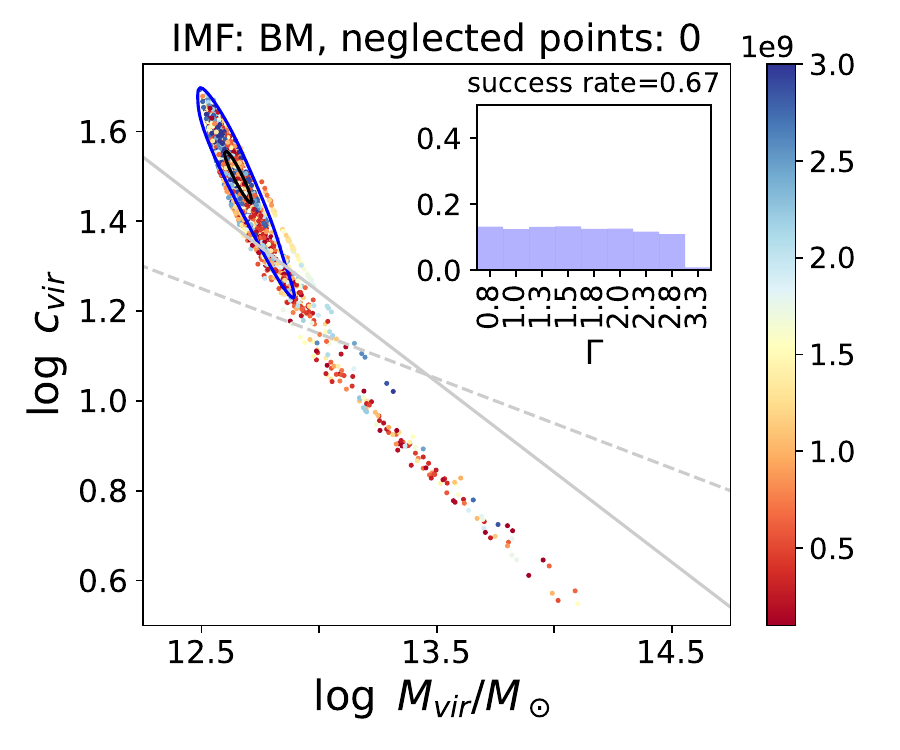}
\includegraphics[scale=0.27]{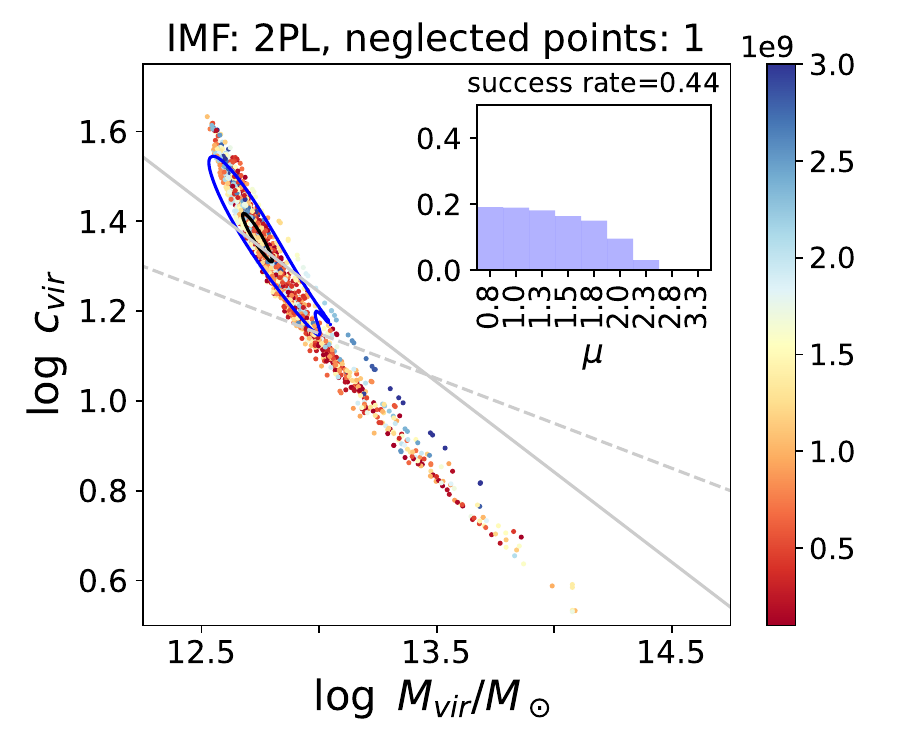}
\includegraphics[scale=0.27]{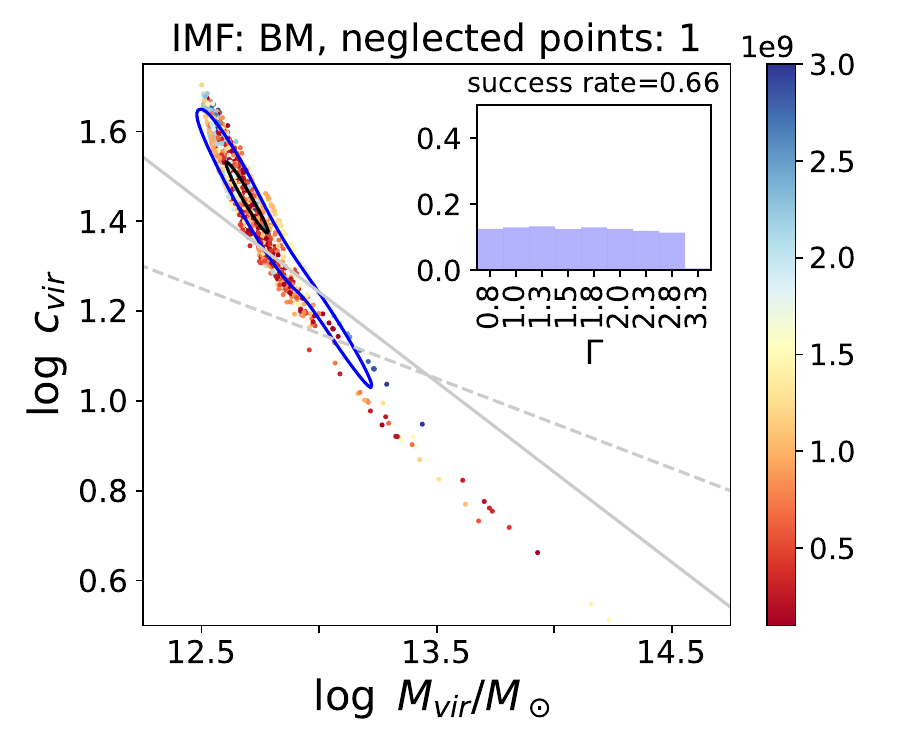}
\includegraphics[scale=0.27]{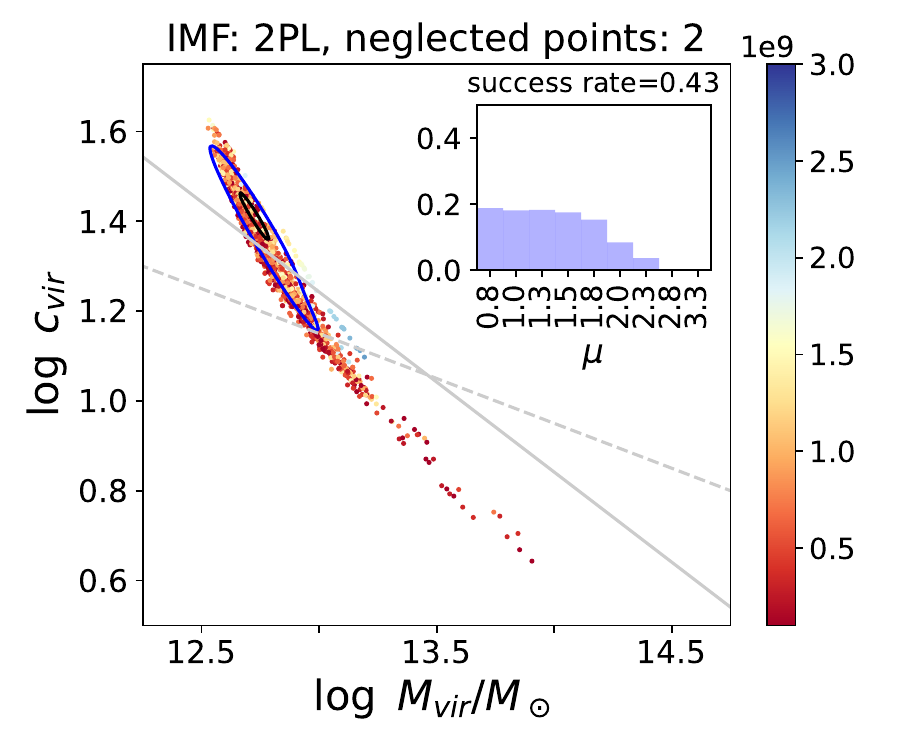}
\includegraphics[scale=0.27]{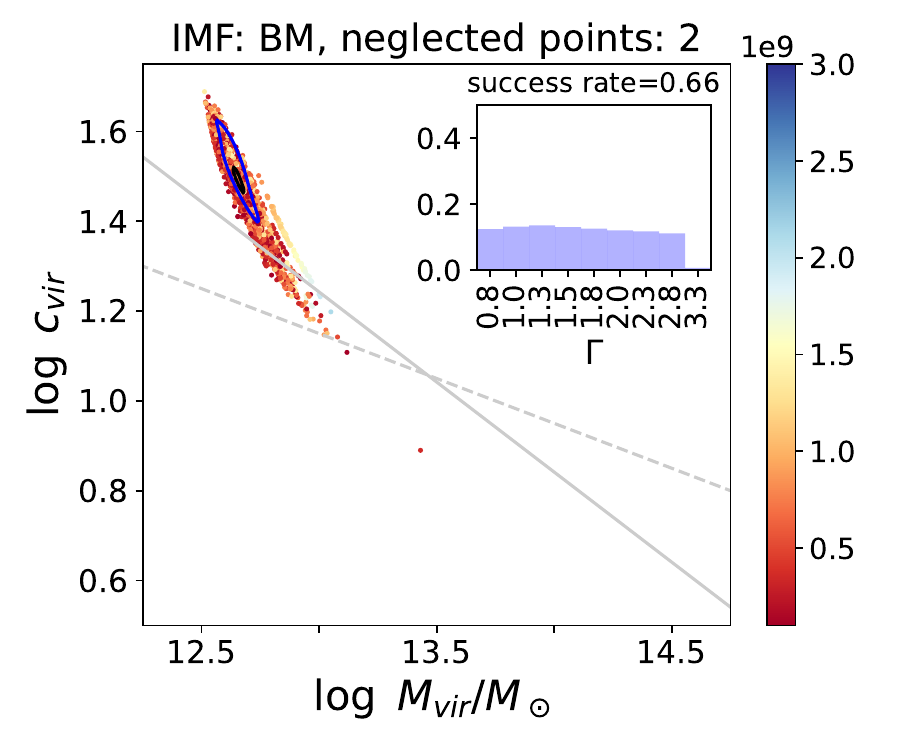}
\includegraphics[scale=0.27]{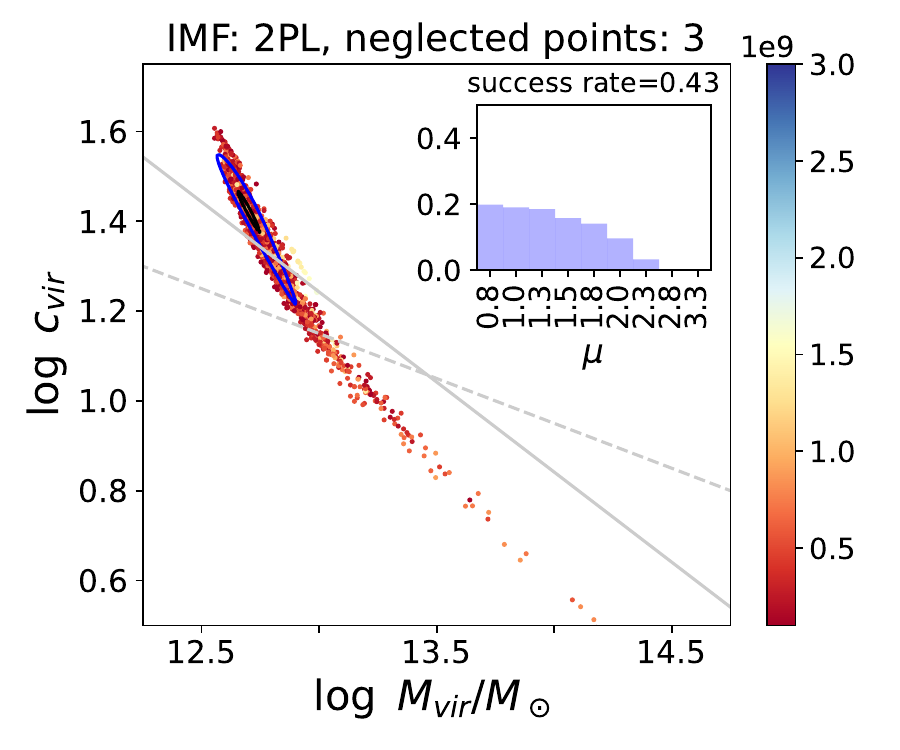}
\includegraphics[scale=0.27]{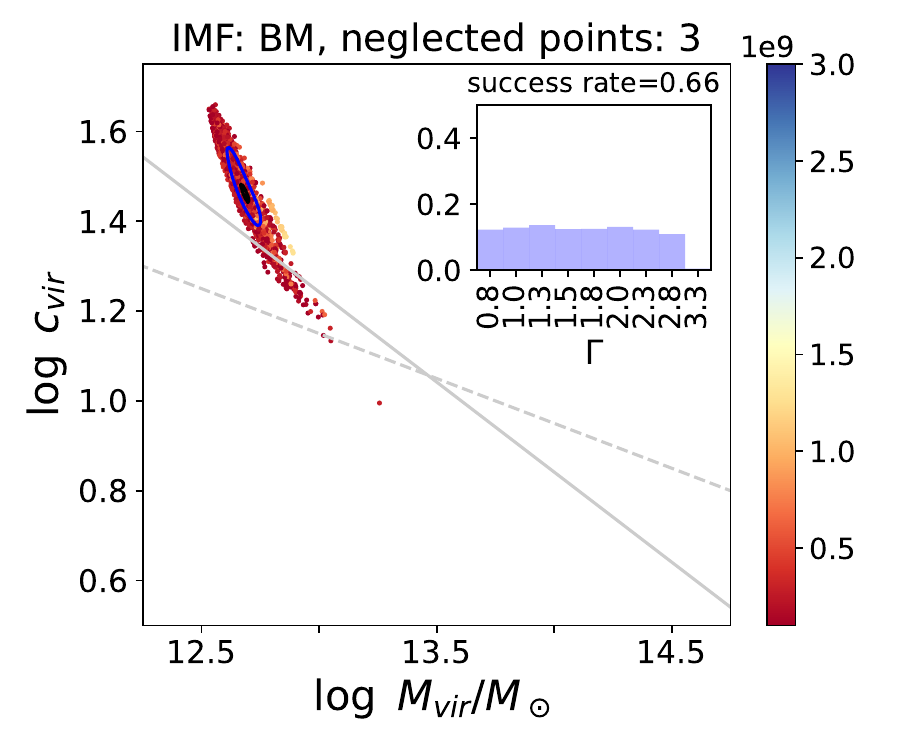}
\caption{Same as in Fig.~\ref{fig:cm1} but with a lower limit added to the
  radial profile. The left (right) column shows the data for a two
  power law (bimodal) IMF. From top to bottom, we show the effect of
  decreasing resolution of the inner radial region by dropping the
  inner zero to three supporting points of the reconstructed mass profiles
  used for the NFW-fit. The colour bar indicates the mean average error
  of the fits in units of $10^9\Msun$.} 
\label{fig:wo}
\end{figure}

Figure~\ref{fig:wo} addresses the potential impact of the spatial
resolution of the central region of the lens on the c-M relation. The
results for the lens J0946 are shown (from top to bottom) when zero to three
inner radial points are neglected in the fitting procedure, as
labelled. The amount of available radial points is derived from a 31x31 grid used in the lens reconstruction method, corresponding to a resolution limit of one-fifteenth of the lens reconstruction radius, as listed in Col.~3 of Table~\ref{tab:lens_data}. By excluding up to three inner radial points from the fit, we assessed both the influence of assuming an NFW profile on the c-M relation and the robustness of the procedure. We present the results for the two choices of stellar IMF:
two power law (left column) and bimodal (right column).

For all lenses as well as the given example of the lens J0946, the
inner part of the mass profile has only a small impact on the cloud of
acceptable fits:
\begin{itemize}
\item The goodness-of-fit is improved, meaning that both the root mean
  square deviation (RMSD) and mean absolute error (MAE) become smaller
  as a consequence of the reduced number of supporting points.
\item The scatter towards low concentration and high virial mass is
  slightly reduced, and the distribution of points in the c-M plane
  becomes more concentrated
\item There are, however, no significant changes in the success rate
  of the fitting procedure nor any significant overall trends in the
  c-M distribution.
\end{itemize} 
We note that the goodness-of-fit and the success rate of the fit are distinct metrics. The success rate refers to the fraction of valid stellar and lens mass combinations that satisfy the convergence criteria of the least-squares fitting routine. These criteria are not met, for instance, when a given combination of total mass and stellar mass profile yields an enclosed mass profile that is non-monotonic or decreases with increasing radius, as enclosed mass profiles must be monotonically increasing. In contrast, the goodness-of-fit is quantified by measures such as the MAE or other deviations between the model and observed data.
We conclude that a well constrained mass profile extending to small
radii ensures a determination of the c-M parameters with small
uncertainties.  Indeed we find for certain systems (especially J2347,
J2303, J1531) -- the ones that are less well constrained by lensing --
that the concentration increases systematically with the number of
neglected points. Others (e.g. J0959, J1525) show a less pronounced
tendency and overlapping as well as widespread density distributions,
which do not permit one to arrive at an unequivocal conclusion.

\begin{figure}
\centering 
\includegraphics[scale=0.25]{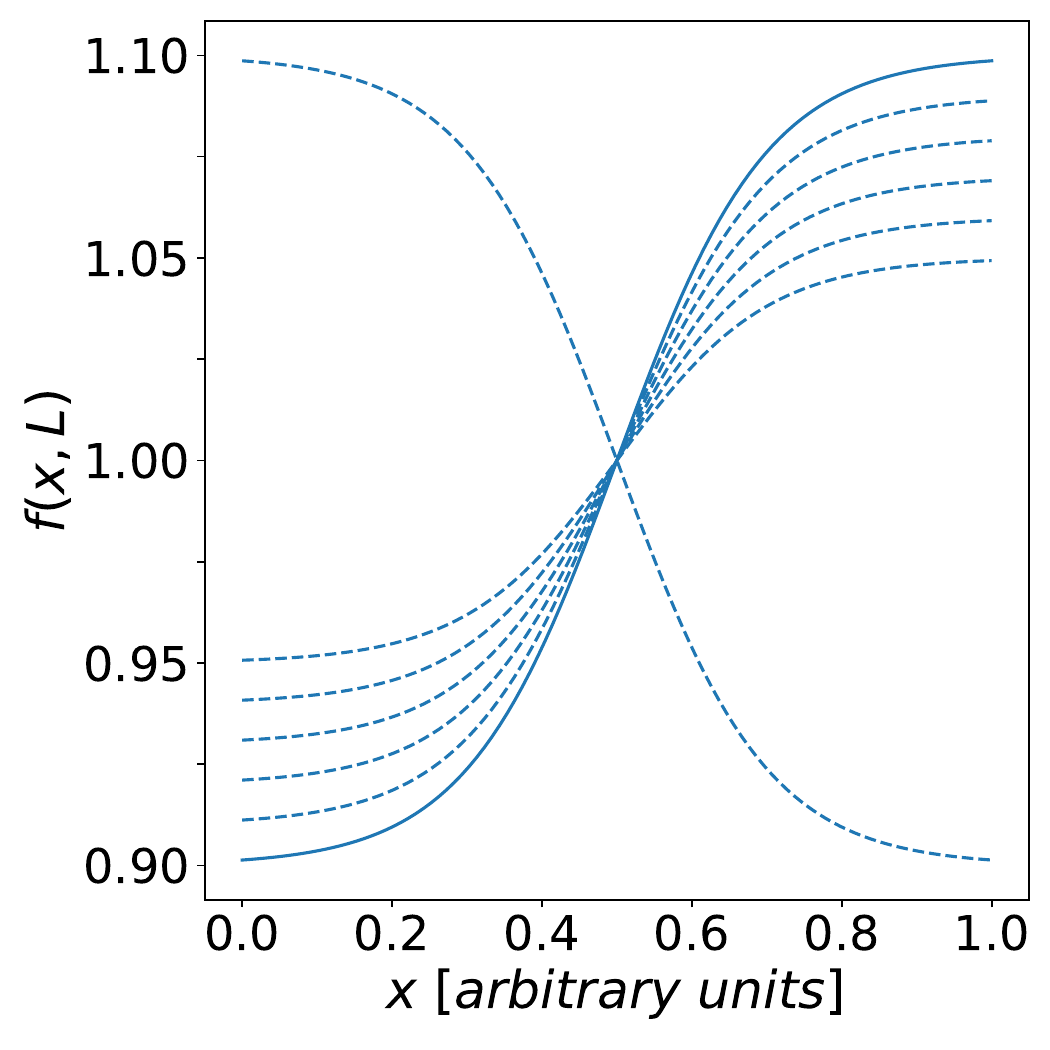} 
\includegraphics[scale=0.25]{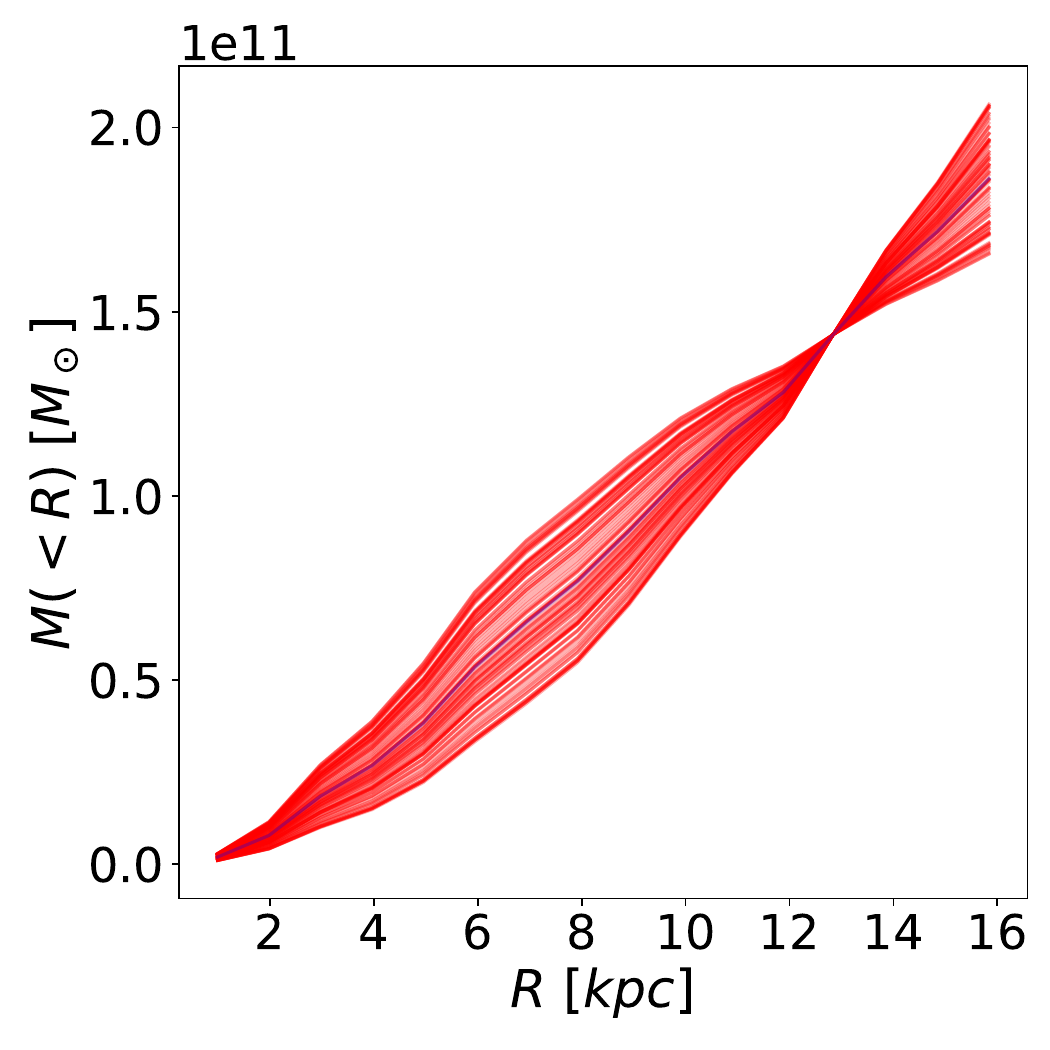}
\includegraphics[scale=0.25]{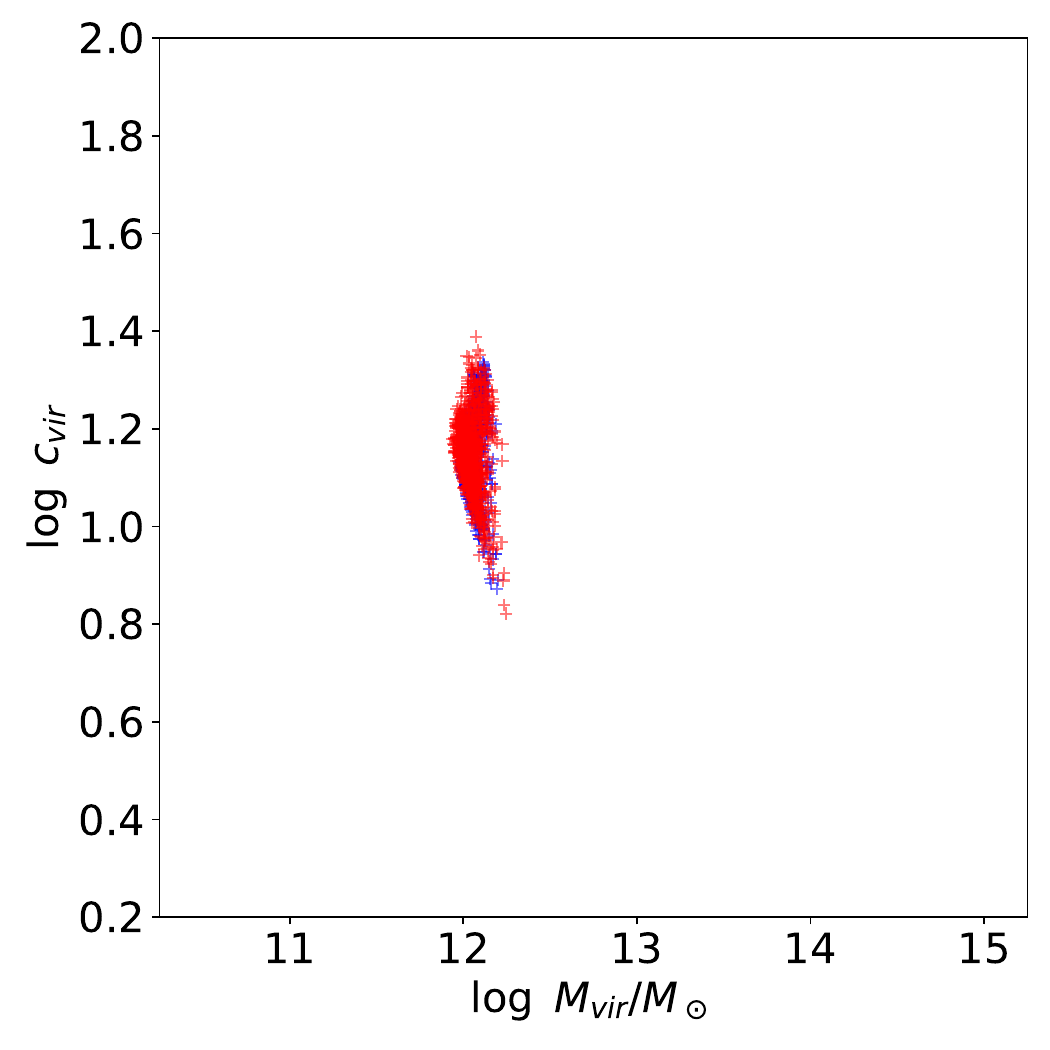}
\includegraphics[scale=0.25]{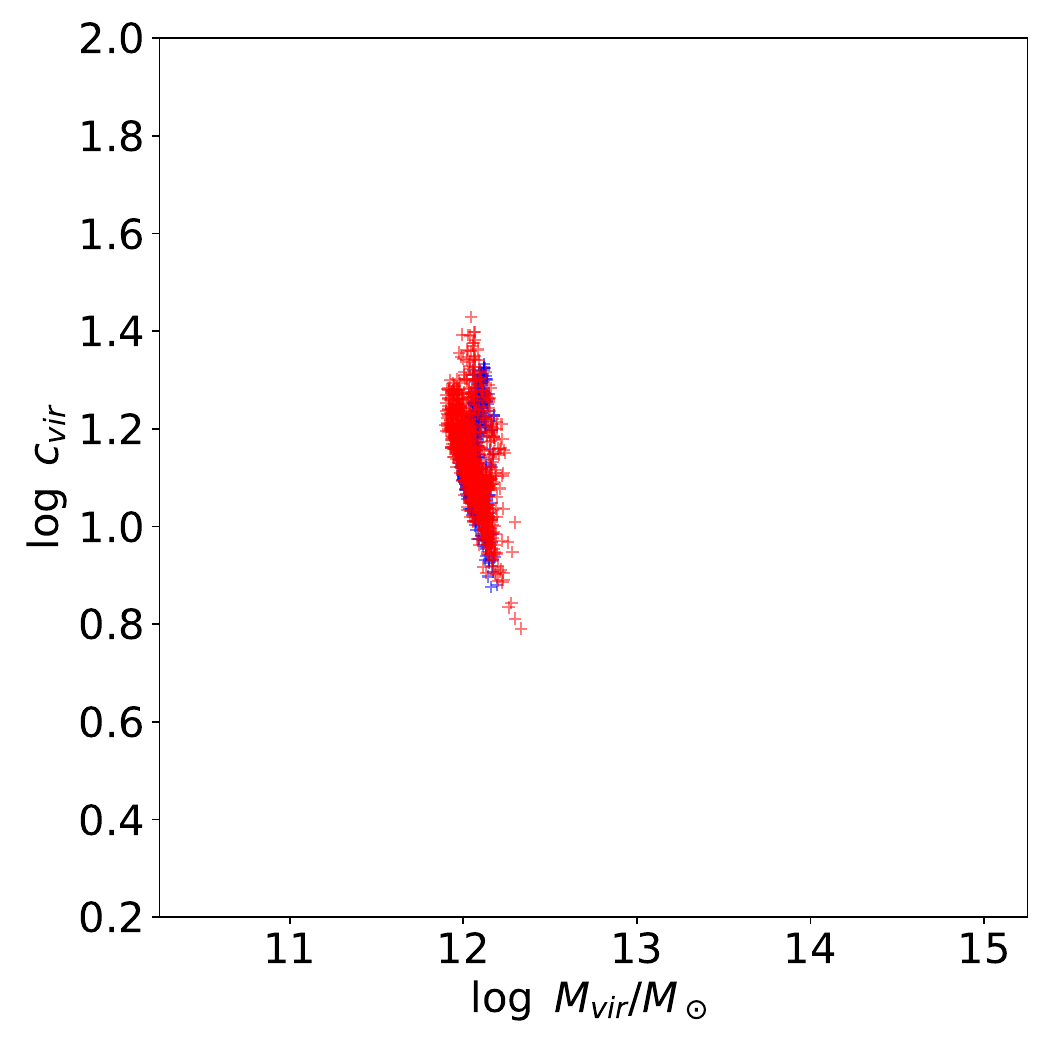} 
\includegraphics[scale=0.25]{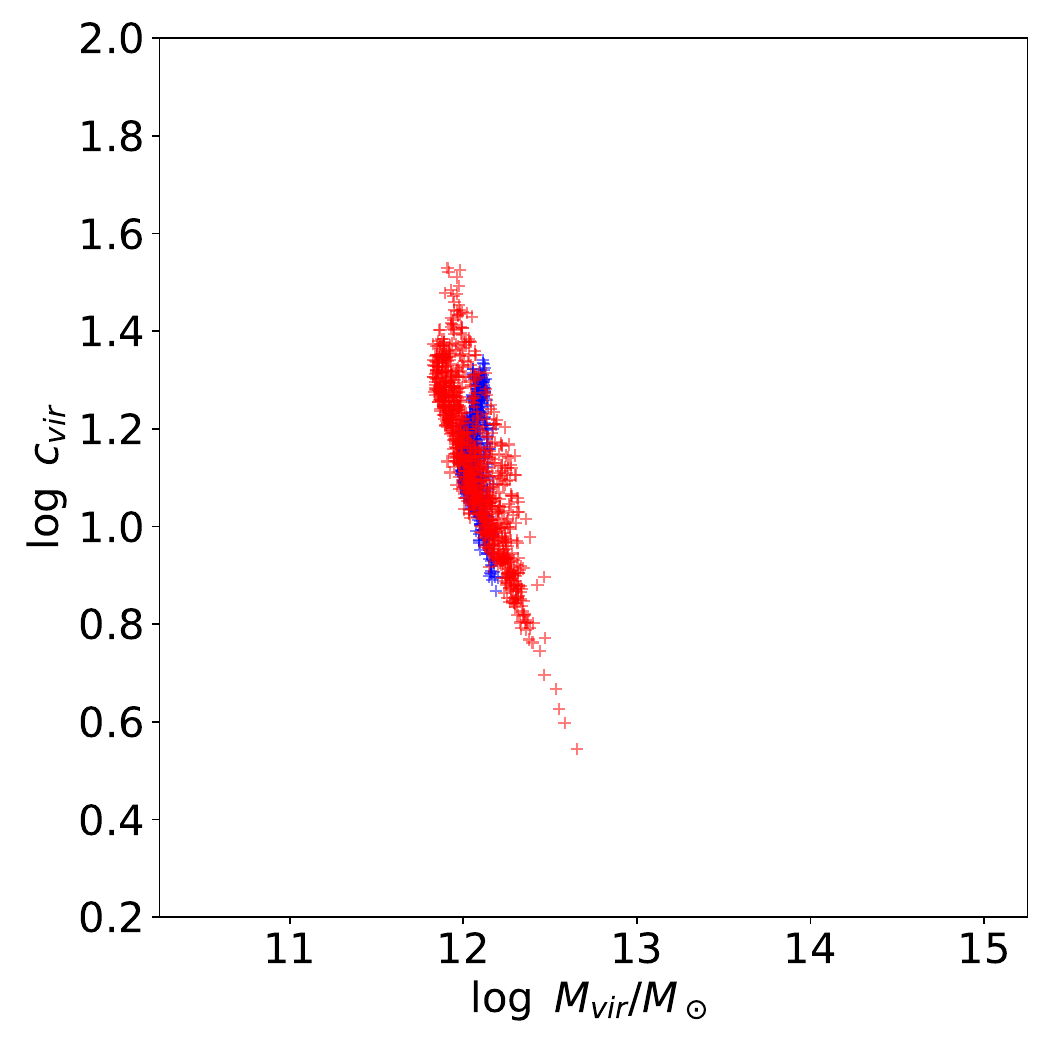} 
\includegraphics[scale=0.25]{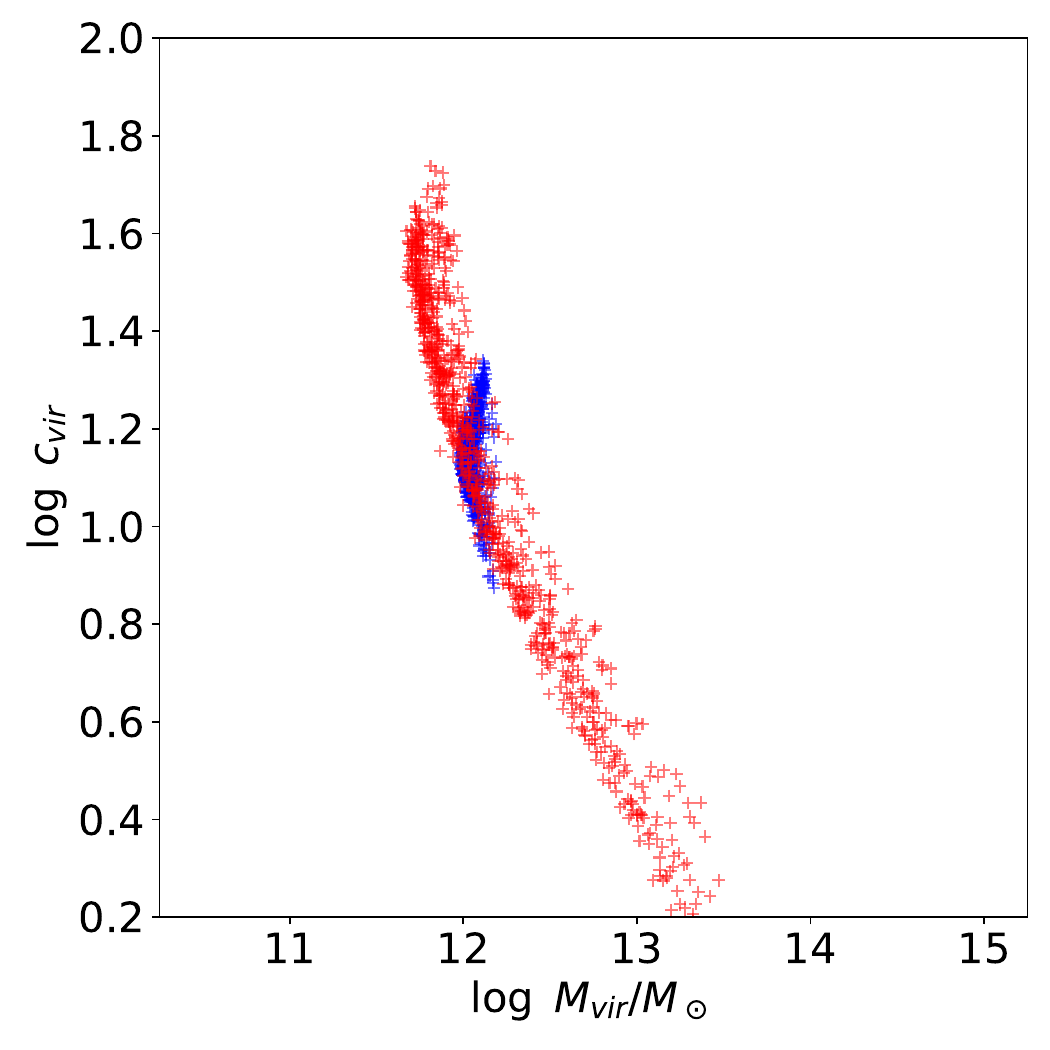} 
\caption{Top left: Distortion function according to
  Eq.~\ref{eq:logit}. The solid line represents the $L=0.1$ case; the
  dashed lines show the cases of $L=0.09$ to $L=-0.1$. Top right: One hundred
  realisations of an enclosed DM mass profile (from J1525) modified by
  random distortion functions with $L$ varying from -0.5 to
  0.5. Middle left to bottom right panel: Changing concentration to
  virial mass relation for 1000 realisation with randomly chosen
  L-values (0.05, middle left; 0.1 middle right; 0.2, bottom left;
  0.5, bottom right) shown in red. The blue point cloud represents the
  fits to the original DM profiles.}
\label{fig:logit}
\end{figure}

Furthermore, in cases with a clear trend between the number of neglected
points and their location on the c-M plane, the effect is independent of the
assumed IMF. We find that the frequency distribution of the free
parameter $\mu$ of the two power-law IMFs (inset plots,
Fig.~\ref{fig:wo}) changes only marginally with the number of
neglected points. The same applies to the free parameter $\Gamma$ of
the bimodal IMFs. There is only a slight tendency for both parameters
to change towards higher values with an increased number of neglected
points, in the sense that we lose a fraction of the $\mu = 0.8$ points
and gain a bit at $\mu = 1.8-2.0$.
    
Hence we conclude that for a few less well-constrained lenses, the
adopted mass reconstruction method may introduce a certain bias in the
distribution of accepted points on the c-M plane. One way to mitigate
or to avoid this issue altogether is by filtering these points with
respect to the goodness-of-fit. The reconstruction method may fit the
data by putting too much mass at the center of galaxies, hence
altering the outcome of the fits. In some cases, however, it seems that
removing a possibly biased innermost point leads to even higher
concentrations. A plausible explanation for this behaviour is related to
the sensitivity of the NFW-fit method to the transition within the
dark matter profile from an initial power law of $r^{-3}$ to the
subsequent $r^{-1}$.
    
When encountering a reconstructed enclosed mass profile that can be
successfully accommodated by a singular power law, the positioning of
the scale radius during the fit tends to converge toward small radii
(i.e. the outermost reconstruction radius). This yields diminished values
for $\rs$ and consequently results in elevated concentration values.
    
As a better understanding of the systematics of lens mass
reconstruction and analytic fits is obtained, it seems worthwhile to investigate in
detail how the output (i.e. the c-M point cloud) is affected and
potentially biased by changes to the steepness of enclosed mass
profiles. We focused especially on transformations that keep the
enclosed mass constant.

\subsection{Enclosed mass profile}
\label{ssec:enclosed}
Another possible bias in the derivation of the c-M relation is
introduced by the radial mass profile.  Gravitational lensing data
feature a well-known butterfly-shaped pattern in the radial enclosed
mass profile, where the uncertainties are smallest at the radial
position $R_{\rm lens}$ of the lensed images (see, e.g., figure~1 in \citealt{EC:10}).
To quantify the effect of systematics in the derivation of the
cumulative mass profile, we applied a distortion function that alters
the steepness of $M(<R)$ inside $R<R_{\rm lens}$ while preserving
$M(<R_{\rm lens})$.

For a halo taken from the EAGLE simulation, we introduced variance in
the enclosed mass profile to mimic the behaviour of lens
reconstruction models being less well constrained. This certainly
applies to a radial region inside the lens radius. The process involves 
differentiating the enclosed mass profiles, namely by calculating the
difference between a supporting point $n$ and its inner neighbour $n-1$
and applying the factor given by the following equation:
\begin{equation}
f(x,L)= 1+L \left( 2 \times {\rm logit}(x|a = 1,k = 10,x_0=0.5 ) -1 \right), \label{eq:logit}
\end{equation}
with the logit function defined as
\begin{equation}
{\rm logit}(x|a,k,x_0)= \frac{1}{1+a \exp\{-k(x-x_0)\}}, 
\end{equation}
followed by cumulating the profile. The parameter $L$ in
Eqn.~\ref{eq:logit} defines the degree of distortion of the
non-cumulative profile as shown in Fig.~\ref{fig:logit} for $L=0.1$,
where the inner part of the mass profile is modified to start at a
value $10\%$ lower than the original mass, and the outer part is
defined to end up with a mass $10\%$ higher than the original
result. By allowing this amount of variance for EAGLE haloes, we
recreated the uncertainty of the ensemble of lens models. As a result,
the cloud of c-M values from acceptable NFW-fits changes from a
tightly constrained region in the c-M plane to a scattered pattern,
elongated along the direction of higher concentration towards lower
mass haloes and lower concentration in higher mass haloes. This
elongated distribution of c-M values has been  found among the observed
lenses if the ensemble profiles are less well constrained. They are
also found in given simulated EAGLE haloes when the range of
projection angles yield very different 2D maps (e.g. due to strong
triaxiality). We note that independent of $L$, the median $c$ and median
$M$ values are close to the real (undistorted) quantities and the
density contours overlap to some degree.

We conclude from this experiment that despite its uncertainties, the median values of the c-M fits deduced by our method should be reliable estimates of the real quantities, given that no other bias affects our lens sample. In the next section, we investigate fitting systematics affecting the parametric concentration by deriving a non-parametric concentration that is unaffected by the underlying assumption of NFW-distributed dark matter.

\subsection{Parametric concentration}
\label{ssec:paramc}
Since $\cvir$ is defined as the virial radius divided by the NFW scale
radius, the definition strongly relies on the profiles at small
radii. Therefore, it seems prudent to contrast these results with an
alternative, more robust non-parametric concentration defined as
$R_{90}/R_{50}$, where $R_X$ is defined as the projected 2D radius
within which the mass enclosed is X percent of the enclosed total mass within two effective radii $M(<2\re)$. For instance,
in \citet{Leier:11} it was found that the $R_{90}/R_{50}$ ratio behaved
differently for the lensing and stellar mass distributions, with
the stellar component being more concentrated in more massive galaxies.
 In our case, we wanted to assess whether $R_{90}/R_{50}$ for the total mass is also found to be different in observed and simulated galaxies. 
This ratio was measured directly from the lens and EAGLE ensembles and independently of
the parametric fits.

\begin{figure}
\centering 
\includegraphics[scale=0.45]{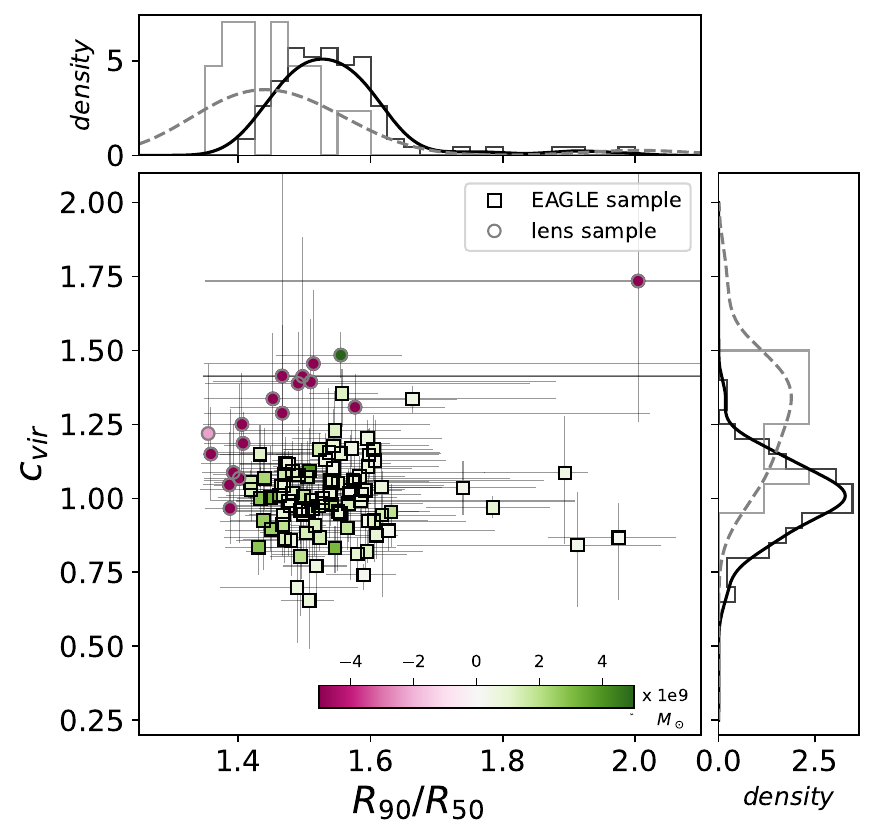} 
\caption{Virial concentration versus non-parametric $R_{90}/R_{50}$. The colours represent median values of the residuals of the NFW-fits. The lens sample (circles) shows mostly large negative residuals (magenta), whereas the EAGLE sample (squares) shows small positive residuals (green). The X- and Y-axis histograms show two populations: kernel densities for the EAGLE sample (solid black line) and the lens sample (dashed grey line). A two-sample KS test for $c_{vir}$ and $R_{90}/R_{50}$ for a significance level of $\alpha=0.001$ indicated that the two samples are drawn from different populations.}
\label{fig:ccrr}
\end{figure}

In Fig.~\ref{fig:ccrr} we show the non-parametric concentrations
($R_{90}/R_{50}$) of the EAGLE (squares) and lens sample (circles) and
compare it to the parametric concentration $\cvir$. We highlight the
goodness-of-fit in terms of median residuals of the NFW-fit by
colour. The lens sample exhibits negative residuals, as the fits are
larger than the data points, especially at small radii. The EAGLE
sample provides small positive residuals. Although the point clouds
overlap slightly, a two-sample Kolmogorov-Smirnov (KS) test for $\cvir$ and $R_{90}/R_{50}$
for a significance level of $\alpha=0.001$ rejects the null
hypothesis. The two samples are thus drawn from different
distributions, meaning that both the parametric and the non-parametric
concentration are distinct when comparing simulated haloes and
lenses. The qualitative results are not affected by excluding singular lenses nor by adding random noise.
However, the error bars are not taken into account for the
KS test. Considering the large uncertainties, this conclusion might be
premature (especially in the presence of bias).

When we investigated the residuals of the NFW-fits, we found that most
lenses exhibit strong negative residuals. In other words, the model fits yield
a smaller mass than the original data -- which is mostly due to
departures from the NFW profile at small radii. The EAGLE haloes in
contrast can be well fitted by NFW profiles with only small positive
residuals.

As lens and EAGLE concentrations are distinct, according to both their
parametric and non-parametric definitions, the NFW-fit itself cannot be
the root cause of the c-M discrepancy. The fact that lenses are less well described by NFW profiles may however 
originate from the models being too concentrated in the inner
radial region, as a consequence of the subtraction of baryon mass or
possibility due to a known tendency of PixeLens models to oversample
highly concentrated models \citep{2012MNRAS.425.3077L}.
However, as we used the NFW-fit exactly for the purpose
of getting rid of this shortcoming, we conclude that our method
produces the expected results and is less affected by the reconstruction bias.
After establishing that the NFW profile assumption plays a
crucial role in our method (i.e. removing the aforementioned bias),
we wanted to study what happens if we choose
different analytic descriptions for the dark matter component.

\subsection{Beyond NFW profiles}
\label{ssec:emp}
Since c-M studies are mostly based on the NFW scale radius definition
of $\rs$ and hence $\cvir$, which we are trying to reconcile with our
findings, assuming different dark matter profiles may seem to be a futile
exercise. However, one way of finding out whether or not an NFW
profile is a well-suited representation of the $\Delta M
(<R)=\Mlens(<R)-\Mstel(<R)$ is by checking different profile fits, the
amount of feasible fits to combinations of $\Mlens(<R)$ and
$\Mstel(<R)$, and their goodness-of-fit.

\begin{figure}
\centering 
\includegraphics[scale=0.55]{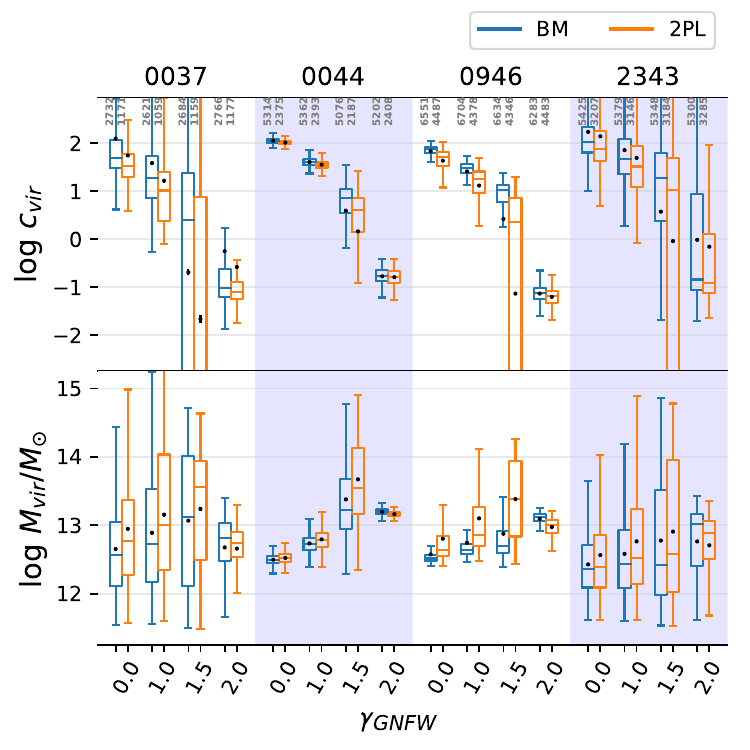} 
\caption{Box plot of concentrations (top panel) and virial masses (bottom panel) plotted against $\gamma_{\rm GNFW}$ for the four lenses J0037, J0044, J0946, and J2343. Mean values and standard errors are shown as black dots with error bars in addition to the box plots showing the interquartile range (25-75\%) and whiskers at the $90\%$ confidence interval. The small vertical numbers show how many of the original 10,000 realisations produced viable fits.}
\label{fig:GNFW}
\end{figure}

In this section, we change the description of the underlying analytic
function, which in turn changes the significance of the scale radius
used in the definition of the concentration $\rvir/\rs$. For this, we
used the projected cuspy NFW profiles as defined in \citet{Keeton_2001}.
This model is a generalised version of the NFW profile, with a density profile of
\begin{equation}
  \rho(r)\propto \left(\frac{r}{r_s}\right)^{-\gamma_{\rm GNFW}}
  \left(1+\frac{r}{r_s}\right)^{\gamma_{\rm GNFW}-3}.
\end{equation}
The cumulative projected mass profiles for $\gamma_{\rm GNFW}=\{0, 1, 1.5, 2\}$ are as follows:
\begin{eqnarray}
\gamma_{\rm GNFW}=0: & M \sim 2\kappa_s r_s \times \left[ 2 \ln \frac{x}{2} + \frac{x^2 + (2-3x^2) \mathcal{F}(x)}{1-x^2} \right] \\
\gamma_{\rm GNFW}=1: & M \sim 4\kappa_s r_s \times \left[ \ln \frac{x}{2} + \mathcal{F}(x) \right] \\
\gamma_{\rm GNFW}=1.5: & M \sim 4\kappa_s r_s x^{3/2} \times \nonumber \\
& \Big[ \frac{2}{3} {}_2F_1[1.5,1.5,2.5,-x] + \Big. \nonumber  \\
&  \left. \int_0^1 dy (y+x)^{3/2} \frac{1 - \sqrt{1-y^2} } {y} \right] \label{eqn:int} \\
\gamma_{\rm GNFW}=2: & M \sim 4\kappa_s r_s x \times \left[ \frac{\pi}{2} + \frac{1}{x} \ln \frac{x}{2} + \frac{1-x^2}{x} \mathcal{F}(x) \right],
\end{eqnarray}
where $x=r/r_s$, ${}_2F_1$ is the hypergeometric function and
\begin{equation}
  \mathcal{F}(x)=
  \left\{\begin{aligned}
  (x^2-1)^{-1/2}\tan^{-1}\sqrt{x^2-1} & \qquad & x>1\\
  (1-x^2)^{-1/2}\tanh^{-1}\sqrt{1-x^2} & \qquad & x<1\\
  1 & \qquad & x=1.
  \end{aligned}
  \right.
\end{equation}
We note that we adopted the convention in which the cusp slope has positive
values. We proceeded to use the
extreme values presented above of $\gamma_{\rm GNFW}$ to study the impact of a
non-standard mass profile on the c-M relation. This is despite the fact that only extreme concentrations allow lens
systems with slopes of 0.5 that can produce multiple images
\citep[see, e.g.,][]{MM:06} and that our sample of observed lenses exhibit
quad configurations that all have very visible lensed images, which already sets boundaries for the mass distribution. By imposing different
slopes in the fit, we essentially changed the impact that data at small
radii have on extrapolated quantities, including the virial radius, and
we also changed the meaning of the scale radius. To determine the
virial radius, we considered the mass enclosed within a sphere of the
profiles presented above. For the different values of $\gamma_{\rm GNFW}$, we obtained
the following enclosed masses:
\begin{eqnarray}
\gamma_{\rm GNFW} = 0: & 4  \pi  \rho_s r_s^3   \Big(\log(1 + x)-\frac{x (2+3 x)}{2 (1+x)^2}\Big) \\
\gamma_{\rm GNFW} = 1: & 4 \pi  \rho_s r_s^3 \Big(\log(1 + x) - \frac{x}{1+x}\Big) \\
\gamma_{\rm GNFW} = 1.5: &  4 \pi \rho_s r_s^3 \frac{2}{3}  x^{3/2}  {}_2F_1[1.5,1.5,2.5,-x] \\
\gamma_{\rm GNFW} = 2: &4 \pi  \rho_s r_s^3 \log(1 + x),
\end{eqnarray}
where $x = r/r_s$
In Fig.~\ref{fig:GNFW}, we show the impact on concentration $\cvir=\rs/\rvir$ and virial mass $\Mvir$ of changing $\gamma_{\rm GNFW}$. We
restricted the plot to four out of 18 lenses, namely J0037, J0044,
J0946, and J2343.  The concentrations of the 10,000 random
combinations of total and stellar mass (i.e. $\Delta M = M_{\rm
  tot}-M_{\rm s}$) exhibit a trend towards lower concentration with
increasing $\gamma_{\rm GNFW}$ (i.e. towards cuspier models). Beyond
0.6\% of $r_{\rm vmax}$ , there is a general consensus that haloes are
denser than those derived from an NFW profile and that they are well
fitted by functions that are steeper and cuspier than NFW at these
radii, as with the GNFW profile with $\gamma_{\rm GNFW}=1.2$ or the
Einasto profile with $\gamma=0.17$ \citep[e.g.][]{MD:10}.  While low
$\gamma_{\rm GNFW}$ profiles show a turnover, which can also be seen
in the lens profiles, high $\gamma_{\rm GNFW}$ profiles lack this
additional constraint. Especially profiles with $\gamma_{\rm GNFW}=2$, which is a
single power law without turnover, the fitted scale radius
automatically goes to large values, and $\cvir$ consequently drops to
almost zero. Profiles with $\gamma_{\rm GNFW}=1.5$ seem to bridge the
$\gamma_{\rm GNFW}=1$ and $2$ cases with a large scatter. As
concentration and virial mass are inversely related and belong to a
one-parameter family (a result from assuming one of the above dark
matter functions), an increasing trend of $Mvir$ with
$\gamma_{\rm GNFW}$ follows naturally.

Changing the IMF does not qualitatively alter this result. 
However, with only a few
exceptions, bimodal IMFs produce a smaller variance, as shown in Fig.~\ref{fig:GNFW}.
As can be seen for J0946, the values move gradually from
non-physically large concentrations of $\log c>1.5$ and virial masses
of $10^{12.5}$ $\Msun$ for $\gamma_{\rm GNFW}=0$ to implausibly small
concentrations of $\log c<-1$ and virial masses at around $10^{13.2}$
$\Msun$. The residual mean standard deviation of the fits shifts only
marginally to larger values. It should be noted that small virial
concentrations with $\gamma_{\rm GNFW}=2$ are a result of the scale
radius remaining unconstrained by the adopted mass profile. For all
other profiles the turnover between two different density slopes are a
function of $\rs$. Consequently, the scale radius drifts gradually
towards higher values in the fitting process.

Using the scale radii of respective GNFW profiles to compute $\cvir$
shows that $\gamma_{\rm GNFW}=0$ and $1$ profiles produce similar
results in terms of goodness-of-fit (RMSD and MAE) while reducing its
absolute value. The choice of GNFW also does not exclude certain
combinations of $\Mlens$ and $\Mstel$, as seen by the number of
realisations producing a valid fit at the top of
Fig.~\ref{fig:GNFW}. While a smaller $\gamma_{\rm GNFW}$ produces a larger
concentration, a larger $\gamma_{\rm GNFW}$ reduces the concentration. We
find, however, no evidence that an NFW is a worse representation of
the dark matter profiles that we produced with our method, but we 
acknowledge the fact that $\gamma_{\rm GNFW}<1.5$ profiles produce
suitable fits and that by extension a real $\gamma_{\rm
  GNFW}=0$ profile falsely fitted by an NFW profile might bias the c-M
results. In the next section we focus on the impact of the IMF
choice on concentration and virial mass in more detail.

\subsection{Stellar initial mass function}
\label{ssec:imf}

\begin{figure}
\centering 
\includegraphics[scale=0.55]{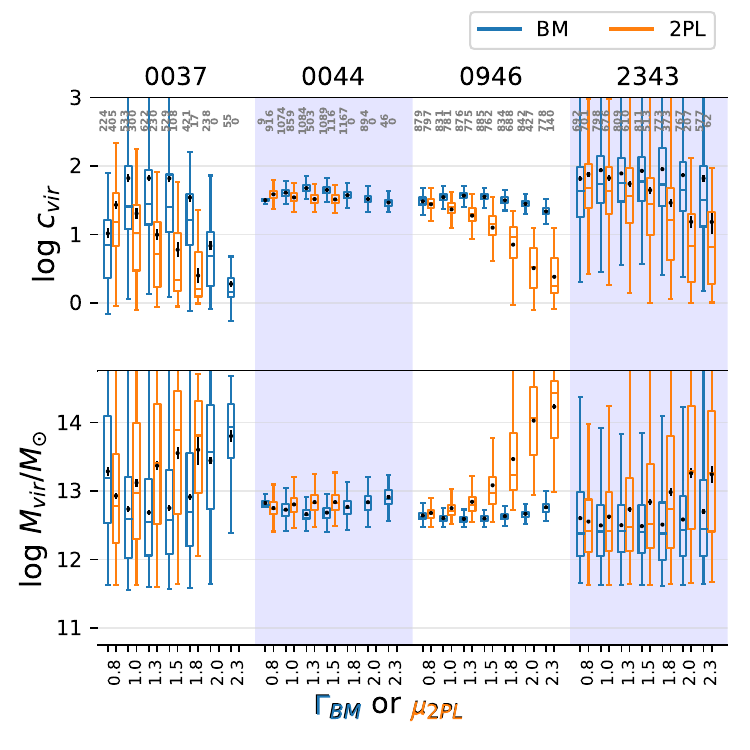}  
\caption{Parameters of c-M as a function of IMF slopes $\Gamma_{BM}$ for a bimodal IMF (blue) and $\mu_{2PL}$ for a two-segment power-law IMF (orange) in case of an NFW $\gamma=1$ fit to the residual enclosed mass profiles. The box plots show the interquartile range plus $90\%$ CI as whiskers. The black dots and error bars show the mean values with standard errors.}
\label{fig:gamma1}
\end{figure}

In Fig.~\ref{fig:gamma1} we show how the c-M parameters change for an NFW fit to the dark matter component if we assume different slopes $\Gamma_{BM}$ for the bimodal IMF defined in \cite{Vazdekis_2003} and $\mu_{2PL}$ for the two-segment power-law IMF of eq.~\ref{eq:2pl},
where $\mathcal{N}$ is a normalisation factor and $\mu$ is a free parameter that controls the fractional contribution in low-mass stars. 
\begin{equation}
    \frac{d\mathcal{N}}{d\log\mathcal{N}} = \mathcal{N} \times \left\{ \begin{array}{ll}
            M^{-\mu} & \quad M \leq \Msun \\
            M^{-1.3} & \quad M > \Msun
        \end{array} \right.
        \label{eq:2pl}
\end{equation}
We note that for the bimodal case, $\Gamma > 1.3$ represents a
more bottom-heavy IMF than Kroupa (2001), whereas for the two-segment
power-law function, $\mu > 1.3$ is more bottom-heavy than Salpeter
(1955). For the two-segment power law, a substantial drop in
concentration and an increase in virial mass can be observed beyond
$\mu=1.5$. Although less pronounced, this trend can be confirmed for
the bimodal IMF, too. We note that for J0037 and J0044, the bottom-heavy
end of the two-segment power law could be ruled out, as none of the
stellar mass maps produced enclosed masses smaller than their respective
total mass. The property to constrain the range of the IMF slope was
studied in \cite{Leier:11} and can be seen in the inset histogram of
Fig.~\ref{fig:cm1}.

\begin{figure}
\centering 
\includegraphics[scale=0.55]{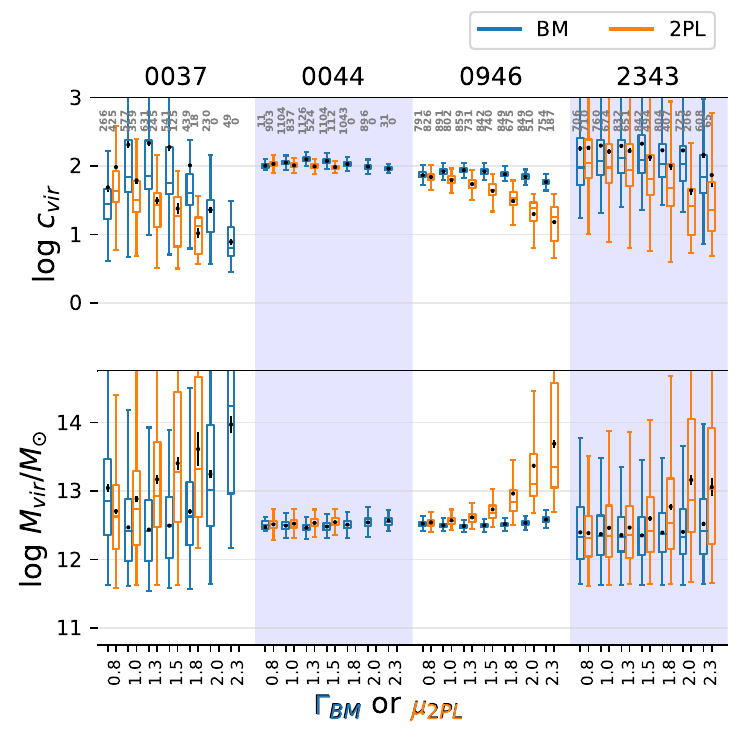} 
\caption{As in Fig.~\ref{fig:gamma1} but for a $\gamma_{\rm NFW}=0$ fit.}
\label{fig:gamma0}
\end{figure}

\begin{figure}
\centering 
\includegraphics[scale=0.55]{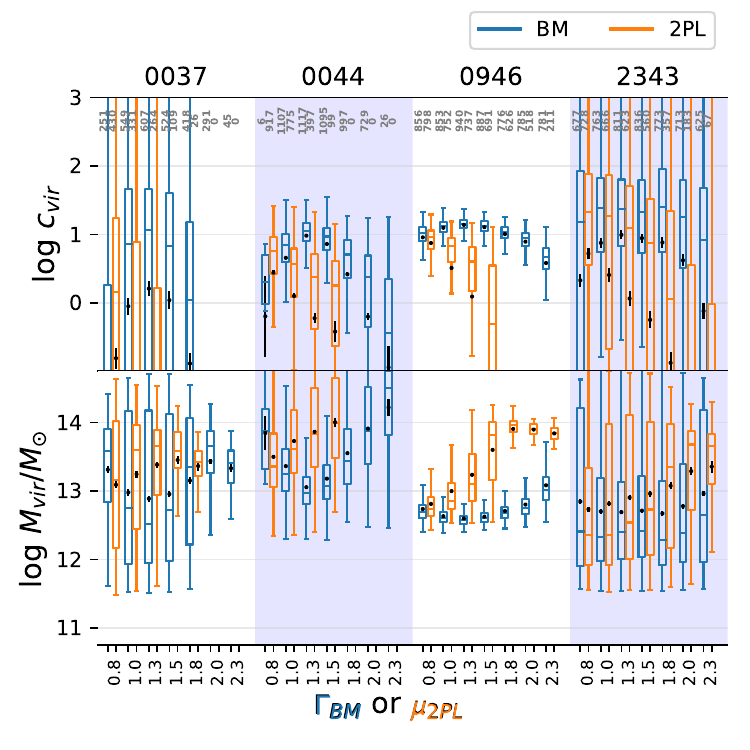} 
\caption{As in Fig.~\ref{fig:gamma1} but for a $\gamma_{\rm NFW}=1.5$ fit.}
\label{fig:gamma15}
\end{figure}

\begin{figure}
\centering 
\includegraphics[scale=0.55]{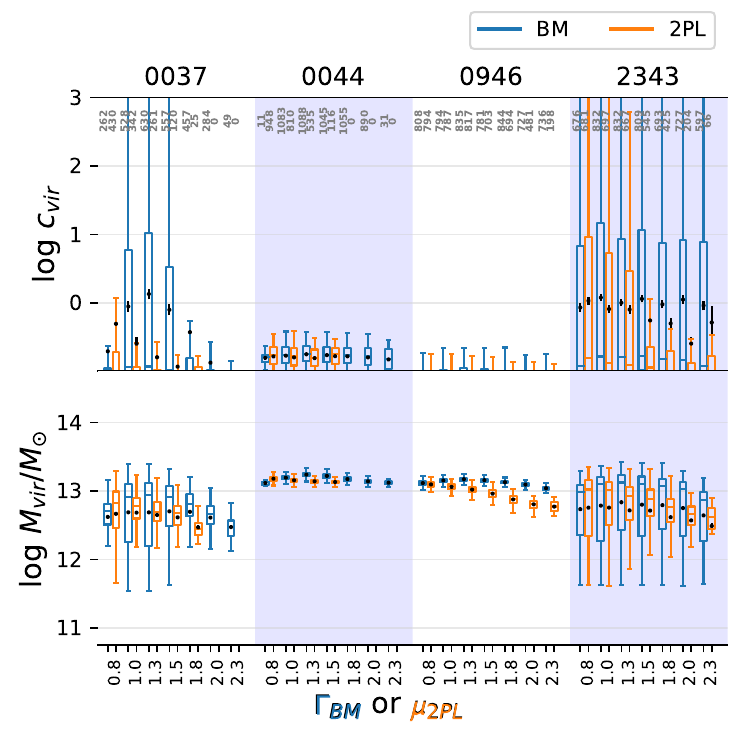} 
\caption{As in Fig.~\ref{fig:gamma1} but for a $\gamma_{\rm NFW}=2$ fit.}
\label{fig:gamma2}
\end{figure}

Through Figs.~\ref{fig:gamma0} to \ref{fig:gamma2}, we reconfirm the results from the previous section and show the systematic impact of the IMF slope on the c-M relation. Significant  differences in the distributions of $c$ and $M$ are visible only if we go to extreme choices for $\Gamma_{\rm BM}$ and $\mu_{2PL}$, respectively.
Only more well-constrained lensing systems, such as $J0946$, are capable of showing the trend with sufficient significance.

\section{Discussion}
\label{sec:disc}

In this paper, we present a series of findings regarding a number of factors
that may influence the conclusions in \citet{LFS22} concerning the offset
on the concentration-mass plane of observed and simulated dark matter
haloes. In particular, we show the following:
\begin{itemize}
\item Via strong lensing, we probed the inner radial region up to the
  scale radius of the targeted DM profiles. The adoption of cuspier profiles
  that can produce slightly better fits causes the scale radius to
  increase faster than the extrapolated virial radius, 
  effectively leading to lower estimates of concentration.
  The case of $\gamma\sim 2$ yields a single power-law distribution without discernible
  turnover and scale radius, causing the fits to reach extreme
  values and hence producing unrealistically low concentrations. Nevertheless,
  we noticed a significant negative (positive) trend between $c$ ($M$)
  and $\gamma_{\rm GNFW}$ when going from $\gamma_{\rm GNFW}=$0.5 to 1.5. The
  uncertainties vary strongly among the lenses in our sample. However,
  changes of $\sim$0.3\,dex in $\log c$ with every 0.5 step in
  $\gamma_{\rm GNFW}$ can be explained.
\item The fact that in some cases GNFW produces better fits suggests a
  strong impact of the inner radial region, which should be considered
  as well when interpreting the results of the NFW
  fits. Moreover, this suggests that the substructure -- which also introduces
  a departure from an NFW profile -- may also induce a higher concentration.
  The inherent limitations of the simulation data lead to better NFW
  fits and thus a lower concentration.
\item Furthermore, we find that changes to the lens mass profile
  affecting its steepness for $r<r_{\rm Ein}$ cause more widespread
  distributions in the c-M plane in terms of both concentration and
  virial mass. This explains the lens-to-lens variation in terms of
  the quality of the available constraints. Less well
  constrained lenses permit a wider range of slopes. Shallower profiles 
  at low radius produce lower concentrations but higher virial
  masses and vice versa.
\item By introducing a non-parametric concentration based on the curve of growth, 
  $R_{90}/R_{50}$, we find tentative evidence that the differences in
  the concentration to virial mass plane presented in \cite{LFS22} can be
  traced back to differences of the dark matter profiles in the inner
  radial region of haloes. The large error bars, however, prevent a
  definitive conclusion.
\end{itemize}


The above results apply for both choices of stellar IMF, though it
should be noted that the 2PL IMF produces systematically
lower concentrations and mostly larger virial masses. We note that this
choice of IMF also consistently produces lower fit success rates, whereas
the results from the BM IMF are less likely to produce unphysical negative mass densities.
\begin{itemize}  
\item A detailed study of IMF slopes showed that bottom-heavy IMFs can be ruled out in
  some lenses, namely $\mu_{2PL}>1.5$ ($1.8$) for J0044 (J0037). This finding holds for all GNFW types.
\item For the 2PL IMF, there is a significant
  inverse trend between concentration and IMF slope. Consequently, the
  virial mass and IMF slope are positively correlated. The decrease in concentration 
  related to the change in IMF slope depends strongly on the lens under study.
  For J0946, we determined a roughly 0.3\,dex change in $\log c$ between 
  $\mu_{2PL}=0.8$ and $1.8$, whereas for J0044, $\log c$ drops by only 
  $10\%$ when going from $\mu_{2PL}=0.8$ to $1.5$.
\item For the bimodal IMF, we find a less pronounced decrease in
  concentrations. We note that the generic definitions of 2PL and BM
  produce a substantial difference in the predicted stellar mass
  with respect to the running parameter ($\Gamma$ or $\mu$). The BM
  definition affects both the low- and high-mass end of the IMF, so
  similar stellar masses could be produced in either case (the top-heavy
  case dominated by remnants and the bottom-heavy case dominated by
  low-mass stars; see, e.g., \citealt{Capp:12}). For the 2PL case,
  the massive end is tied to the Salpeter slope, so only changing $\mu$ monotonically increases the derived stellar mass, leading to a more
  drastic variation of the halo fits. 
  In any case, the success rate of DM-stellar mass
  combinations drops drastically towards the bottom-heavy end of IMF
  slopes for both IMFs.
\end{itemize}

\section{Conclusion}
\label{sec:con}
Our study demonstrates that variations in the c-M relation between simulated and observed dark matter haloes can largely be attributed to differences in the inner mass profiles and the assumed dark matter density profile. By exploring non-standard dark matter models and adopting a more flexible slope for the GNFW profile, we reconciled much of the tension between simulations and observations. Additionally, introducing a non-parametric concentration based on the curve of growth ($R_{90}/R_{50}$) provided tentative evidence that discrepancies in the c-M plane arise from variations in the inner dark matter profiles, though large uncertainties limit definitive conclusions. The choice of IMF also plays a crucial role, with bottom-heavy IMFs systematically producing lower concentrations and impacting fit quality. While our findings highlight important trends, further refinement in both observational constraints and modelling techniques is needed, especially in the inner radial regions.

\begin{acknowledgements}
The research of DL is part of the project GLENCO, funded under the
European Seventh Framework Programme, Ideas, Grant Agreement
no. 259349. IF is partially supported by grant PID2019- 104788GB-I00
from the Spanish Ministry of Science, Innovation and Universities
(MCIU).
\end{acknowledgements}

\section*{Data availability}
This sample is extracted from publicly available data from the CASTLES
sample (Falco et al. 2001) and the EAGLE simulations (Schaye et
al. 2015). The combined final data set is available upon reasonable
request.

\bibliographystyle{aa}
\bibliography{references}

\end{document}